\documentstyle[prl,aps]{revtex}

\begin{document}

\title{Nearly Singular Magnetic Fluctuations in 
the Normal State of a High-T$_{\bf c}$ Superconductor}  
\author{G. Aeppli$^{1,2}$,\ T. E. Mason$^{2,3}$,\ S. M.
Hayden$^4$,\ H. A. Mook$^5$, and J. Kulda$^6$}

\address{$^1$ NEC Research Institute, 4 Independence Way, Princeton,
NJ 08540, USA\\
$^2$ Ris\o\ National Laboratory, 4000 Roskilde, Denmark\\
$^3$ Dept. of Physics, University of Toronto, Toronto, Canada
M5S 1A7\\
$^4$ Dept. of Physics, University of Bristol, Bristol BS8 1TL,
United Kingdom\\
$^5$ Oak Ridge National Laboratory, Oak Ridge, TN 37831, U.S.A.\\
$^6$ Institut Laue-Langevin, BP 156X, Grenoble Cedex, France}

\date{Science, {\bf 278}, 21 November 1997, p. 1432-1435}
\maketitle

\begin{abstract}

Polarized and unpolarized neutron scattering was used to measure the
wave vector- and frequency-dependent magnetic fluctuations in the
normal state (from the superconducting transition temperature, 
T$_{\rm c} =35$, up to 350 K) of single
crystals of La$_{1.86}$Sr$_{ 0.14}$CuO$_4$.  The peaks which dominate
the fluctuations have amplitudes that decrease as $T^{-2}$ and widths
that increase in proportion to the thermal energy, k$_{\rm B}T$ 
(where k$_{\rm B}$ is Boltzmann's constant), and energy transfer added
in quadrature.  The nearly singular fluctuations are consistent with a
nearby quantum critical point.

\end{abstract}

\pacs{PACS numbers:  74.72.Dn, 74.25.Ha, 61.12.Ex}


The normal state of the metallic cuprates is as unusual as
their superconductivity.
For example, the electrical resistivity of samples with
optimal superconducting properties is linear in temperature
(T) from above 1000 K to ${\rm T}_{\rm c}$\cite{ref1}.
Correspondingly, infrared reflectivity reveals charge fluctuations with
a characteristic energy scale 
that is proportional only to T\cite{ref1,ref2}.
Furthermore, the effective number of charge carriers, as measured
with the classic Hall effect, is strangely
T-dependent. Even so, the Hall angle, a measure of the
deflection of carriers in the material by an external magnetic field,
follows a ${\rm T}^{-2}$ law \cite{ref3}. 
Thus, the metallic charge carriers in the doped cuprates exhibit peculiar 
but actually quite simple properties \cite{ref17} in
the normal state.
Also, these properties do not vary much between the different high-T$_{c}$
families.  

Electrons carry spin as well as charge, so it 
is reasonable to ask whether the normal state magnetic properties,
derived from the spins, are as simple and universal as those derived
from the charges. Experiments to probe the spins
include classical magnetic susceptometry,
where the magnetization in response to a homogeneous
external magnetic field is measured, and resonance
experiments, where nuclear dipole and quadrupolar relaxation is
used to monitor the electron spins. The spin-sensitive measurements
yield more complex and less universal results than those sensitive to
charge, and indeed do not seem obviously related to the 
frequency-dependent conductivity, $\sigma(\omega,T)$, 
probed in electrical, microwave 
and optical
experiments. In particular, there is little evidence for magnetic 
behavior which is as nearly singular, in the sense of diverging (for
$T \rightarrow 0$)
amplitudes, time constants, or length scales, as the behavior of
$\sigma(\omega,T)$. 

We report nearly singular
behavior of the magnetic fluctuations in the simplest of
high-T$_{c}$ materials, namely, the  compound La$_{2-x}$Sr$_{x}$CuO$_{4}$ whose
fundamental building blocks are single CuO$_2$ layers, as determined
by inelastic magnetic neutron
scattering.  A beam of mono-energetic neutrons is
first prepared and then scattered from the sample,
and the outgoing neutrons are labeled according to their energies and
directions to establish an angle and energy-dependent scattering
probability, or cross-section, 
$d^2\sigma/d\omega d\Omega$.  Because
the neutron spin and the electron spins in the sample interact through
magnetic dipole coupling, the cross-section is directly proportional
to the magnetic structure function, $S(\bf{Q},\omega)$, 
the Fourier transform of the space- and
time-dependent two-spin correlation function.  The momentum and energy
transfers $\bf{Q}$ and $\omega$ are simply the differences between the
momenta and energies of the in-going and outgoing neutrons,
respectively. Via the fluctuation-dissipation theorem,
$S(\bf{Q},\omega)$ is in turn proportional to the imaginary part,
$\chi^{\prime \prime}(\bf{Q},\omega)$, of the generalized linear magnetic response
$\chi (\bf{Q},\omega)$. The bulk susceptibility measured using a
magnetometer is the long-wavelength, small-wavenumber, (${\bf Q}
\rightarrow 0$), limit of $\chi^{\prime}({\bf Q},\omega=0)$, while the nuclear
resonance techniques are averages of $\chi''({\bf Q},\omega\sim 0)$
over momenta $\bf{Q}$ which are of order inverse interatomic spacings.

Figure \ref{schematics} {\bf A} is a schematic phase diagram for
La$_{2-x}$Sr$_{x}$CuO$_{4}$ as a function of $T$, hole
doping ($x$), and pressure ($y$). Holes and pressure are generally
introduced chemically, most notably through substitution of Sr$^{2+}$ and
Nd$^{3+}$ ions, respectively, for the La$^{3+}$ ions in
La$_2$CuO$_4$\cite{buchner,tranquada}.  Possible magnetic ground
states range from simple antiferromagnetic (AF for $x \sim 0$) to a
long-period spin density wave with strong coupling to the underlying
lattice (shown as a gray "mountain" for $x \sim 0.1$ in
Fig. \ref{schematics} {\bf A}).  Unit cell doubling, where the spin on each
Cu$^{2+}$ ion is antiparallel to those on its nearest neighbors displaced by
vectors $(0,\pm a_{o})$ and $(\pm a_{o},0)$ in the (nearly) square CuO$_2 $ planes,
characterizes the simple AF state\cite{ref9}; the lattice constant,
$a_{o}=3.8$ \AA\ .  The associated magnetic
Bragg peaks, observed by neutron scattering, occur at reciprocal lattice vectors
$\bf{Q}$ of the form $(n\pi,m\pi)$, where $n$ and $m$ are both odd
integers; the axes of the reciprocal lattice coordinate system are parallel to those of
the underlying square lattice in real space.  Substitution of Sr for La 
introduces holes into the CuO$_2$
planes, and initially replaces the AF phase by a
magnetic (spin) glass phase. It is for this non-superconducting
composition regime, where the magnetic signals are strong and for
which large single crystals have long been available, that the most
detailed $T$-dependent magnetic neutron scattering studies
have been performed\cite{ref7}. With further increases in Sr content,
the magnetic glass phase disappears and superconductivity emerges. At
the same time, the commensurate peak derived from the order and
fluctuations in the non-superconducting sample splits into four
incommensurate peaks, as indicated in Fig.
\ref{schematics} {\bf B}\cite{ref8}. 
These peaks are characterized by a position, an amplitude, and a
width. Previous work
\cite{ref8} describes how the peak positions vary with composition at low 
temperatures. The new contribution of the present report is to follow
the red trajectory in Fig.
\ref{schematics}
and so obtain the $T$ and $\omega$ dependence of the
amplitude and width, which represent the maximum magnetic response and
inverse magnetic coherence length, respectively.

The La$_{1.86}$Sr$_{0.14}$CuO$_4$ crystals used here
are the same as those
used in our previous determination\cite{ref10} of 
$\chi ^ {\prime \prime} ({\bf Q}, \omega )$ around and below the
superconducting transition at $T_{\rm c}=35$~K.  Unpolarized 
measurements, where the spins of the in-going and outgoing 
neutrons are unspecified, were carried out using the TAS6 
spectrometer of the
Ris\o\ DR3 reactor in the same configuration employed before\cite{ref10}.  
We  performed measurements with fully polarized
in-going and outgoing beams using the IN20 instrument at the Institut 
Laue-Langevin in Grenoble, France.

Our surveys of ${\bf Q}- \omega$ space at various
$T$s is summarized in Fig. \ref{survey}.  Frame {\bf A} shows scans 
along the solid red line in Fig. \ref{schematics} {\bf B}
through the incommensurate
peaks at $[ \pi ( 1 - \delta ), \pi ]$ and 
$[\pi , \pi (1 + \delta ) ]$ for energy transfer $\hbar \omega$
fixed at 6.1 meV (Fig. \ref{survey} {\bf A}).
We have checked that the peaks are of purely magnetic 
origin by using polarized neutrons (Fig. \ref{survey} {\bf B}). 
The spin-flip (SF) channel
contains background plus magnetic scattering whereas the
non spin-flip (NSF) channel contains background plus phonon
scattering.  The incommensurate response only occurs in the
SF channel, confirming that it is derived from the electron spins.  

The most important result in Fig. \ref{survey} {\bf A} is that the sharp peaks at
80 K broaden to nearly merge at 297 K, an effect also illustrated in
Fig. \ref{survey} {\bf C} and {\bf D}, which illustrate the Q- and $\omega -$
dependence of $\chi ^ {\prime \prime} ({\bf Q}, \omega )$
determined by
using the fluctuation dissipation theorem,
$\chi ^ {\prime \prime} (\bf{Q}, \omega )$ 
$(n ( \omega ) + 1)~=~S ({\bf Q}, \omega )$, where $n ( \omega ) + 1=
1/(1-e^{-\hbar \omega / k_B T})$ ($\hbar$ is Planck's constant divided
by $2 \pi$ and k$_B$ is Boltzmann's constant).
The magnetic structure function S({\bf Q}, $\omega$) is simply the
scattering near the incommensurate peaks measured along the
solid red line in Fig. \ref{schematics} {\bf B} and indicated by the 
filled symbols in {\bf A}, minus the background indicated by the open symbols.
Comparison of Fig. \ref{survey} {\bf C} and \ref{survey} {\bf D} 
shows that warming from T$_{\rm c}$
(35 K) to 297 K  gives a much smaller
$\chi ^{\prime \prime} ({\bf Q}, \omega )$, and eliminates clear
incommensurate peaks at all energies $\hbar \omega$ probed. At intermediate
$T$s there is more modest broadening of the magnetic
peaks at low $\hbar \omega$, as well as an intensity reduction
which is much more pronounced at low than at high $\hbar \omega$.

Figure 3 {\bf A} summarizes the dramatic evolution of the incommensurate peak
amplitudes with $T > T_{c}$.  Both unpolarized and polarized
beam data are shown, and their consistency confirms the  background
subtraction procedure used in the faster unpolarized beam 
measurements.  In addition, full polarization analysis\cite{moon} 
of the intensity 
at the position indicated by a blue dot in Fig. \ref{schematics} {\bf B}
confirms that the increase with $T$ seen in the background shown in
Fig. \ref{survey} {\ A} is of non-magnetic origin.

We have so far given a qualitative survey of our data, which are
astonishing because they display much greater temperature-sensitivity
above the superconducting transition than any other magnetic neutron scattering data collected so far
for a cuprate with composition nearly optimal for superconductivity.
To describe more precisely the singular behaviors of the amplitude and
widths of the incommensurate peaks, we must take into
account the finite resolution of our instrument.  We have consequently
parametrized our data at each $\hbar \omega$ and T
in terms of the convolution of the instrumental resolution and
the general form\cite{ref11},
\begin{equation}
S({\bf Q}, \omega ) =
{[{\rm n}( \omega ) + 1 ] \chi _ {\rm P} ^ {\prime \prime} 
( \omega , {\rm T})  \kappa ^ 4 ( \omega , {\rm T}) \over
[ \kappa ^ 2 ( \omega , {\rm T}) + {\rm R}({\bf Q})] ^ 2} \ .
\end{equation}
\noindent R({\bf Q}) is a function, with the full symmetry of the
reciprocal lattice and dimensions of $|{\bf Q}|^2$, which is
everywhere positive except at zeroes coinciding with the
incommensurate peak positions.  From this definition, it follows that
$\chi _ {\rm P} ^ {\prime \prime} ( \omega , {\rm T})$ (in absolute
units via a standard phonon-based calibration\cite{ref12}) is the peak
susceptibility and $\kappa ( \omega , {\rm T})$ is an inverse length scale
measuring the sharpness of the peaks.  To perform fits, we have
expanded R({\bf Q}) near ($\pi , \pi$) to lowest order in $q_x$ and $q_y$, 
the components of {\bf Q} relative
to $( \pi , \pi )$,

\begin{equation}
{\rm R}({\bf Q}) = {[(q_x -q_y)^2-(\pi\delta)^2]^2
+[(q_x+q_y)^2-(\pi\delta)^2]^2\over
2 (2 {\rm a} _{0} \pi \delta ) ^ 2 }\ .
\end{equation}

Because all of our data show features at
the incommensurate positions at which the low-T and $- \omega$ data
are peaked, we simply fix $\delta$ at its low-T and $- \omega$ value
of 0.245.  The solid lines in Fig. 1 {\bf A} and {\bf B} correspond to
Eq. 1 convolved with the instrumental resolution with parameters
$\kappa ( \omega , T)$ and $\chi _ {\rm P} ^ {\prime
\prime} ( \omega , T)$ chosen to obtain the best fit; that
the data and fits are indistinguishable attests to the adequacy of
Eq. 1 as a description of our measurements.

We discuss first the peak amplitudes $\chi _ {\rm P} ^ {\prime \prime}
( \omega , {\rm T})$, shown for three temperatures as a function of 
$\omega$ in Fig. \ref{survey} {\bf E}.  In agreement our earlier work\cite{ref8},
when these amplitudes are assembled to produce spectra as a function
of $\omega$ for fixed $T$s, there is no statistically
significant evidence for a spin gap or even a pseudogap at any ${\rm
T} \geq T_c$.  Furthermore, only for the lowest
$T$ (35 K), is there an identifiable energy scale below 15 meV.
For 35K, the scale is the $\sim 7$ meV energy transfer beyond which
the peak spectrum flattens out.  Otherwise, all of the data are in the
low-$\omega$ regime where $\chi _ {\rm P} ^ {\prime \prime} ( \omega ,
{\rm T})$ is proportional to $\omega$.  This means that at each ${\rm
T} \geq 85$ K our measurements are characterized by a single
amplitude parameter, namely $\chi _ {\rm P} ^ {\prime \prime} ( \omega
, {\rm T})/ \omega$.  Even for 35~K $\leq T \leq$\ 85~K, this is
true for $\hbar \omega <$\ 5~meV.  We consequently shift our attention
to the detailed T-dependence, shown in Fig. \ref{tdep} {\bf B},
 of the low-frequency limit of
$\chi _ {\rm P} ^ {\prime \prime} (
\omega , T)/ \omega$.

The remarkable result is that the peak amplitude, after correction for
resolution broadening effects, changes by two orders of magnitude over
the one order of magnitude rise in temperature from $T_c = 35$ K.
Indeed, a ${\rm T} ^ {- \alpha}$ law with $\alpha = 1.94 \pm 0.06
\cong 2$ describes the decrease of $\chi _ {\rm P} ^ {\prime \prime} (
\omega , {\rm T})/ \omega$ with increasing T, indicating a divergence
in the $T \rightarrow 0$ limit that is interrupted by the
superconducting transition.

We turn now to how the inverse length $\kappa ( \omega , {\rm T} )$
depends on T and $\omega$. The behavior, shown in Fig. 3 {\bf C} appears
complicated, apart from the fact that raising either T or $\omega$
increases $\kappa ( \omega , {\rm T})$. However, closer inspection
reveals that similar increases are associated with frequencies
$\omega$ and temperatures T where $\rm{k_B T \approx \hbar\omega}$.
Fig. 4, which shows $\kappa ( \omega , {\rm T})$ plotted against
$\sqrt {{\rm T} ^ 2 + ( \hbar \omega / {\rm k} _ {\rm B} ) ^ 2}$,
makes the interchangeablitity of temperature and frequency
obvious.  Here, the $\kappa ( \omega , {\rm T})$ values
for different $\omega$'s cluster near a single line with inverse slope
2000 (\AA K)$^{-1}$ $\cong {1 \over 3} {{\rm Ja}_0 / {\rm k}_{\rm
B}}$, where J is the exchange constant of pure La$_2$CuO$_
4$\cite{ref6} Correspondingly, the solid curve in the figure for
which:

\begin{equation} 
\kappa ^ 2 = \kappa _ o ^ 2  + 
a _ o ^ {-2} [({\rm k} _ {\rm B} {\rm T}/{\rm E} _ {\rm T} ) ^ 2 
+ 
( \hbar \omega /{\rm E} _ \omega ) ^ 2 ] ^ {1/Z}\  ,
\end{equation}

\noindent $Z = 1$, $\kappa _ 0 = 0.034$ \AA $ ^{-1}$,
and ${\rm E} _ {\rm T} = {\rm E} _ \omega = 47 {\rm meV} \cong {1
\over 3} {\rm J}$,
gives a good description of the data. 

As a classical spin system approaches a magnetic phase transition, the
magnetic susceptibility and correlation length typically diverge.  We
have discovered that the normal state magnetic response of La$_{
1.86}$Sr$_{ 0.14}$CuO$_ 4$ is characterized by nearly diverging
amplitude and length scales.
Thus, we are near to a low-$T$ or zero-$T$ phase
transition. The latter is commonly referred to as a quantum critical
point(QCP)\cite{ref14,ref15}, which occurs at T=0 and $\hbar \omega=0$ in a
phase space labeled by $T$, $\hbar\omega$, and a quantum fluctuation
parameter $\alpha$.  The lower inset of Fig. 4 shows such a phase space
where the solid circle marks the QCP.  As for ordinary
critical points, the parameter defining the state of the system
anywhere in the three-dimensional (3D) phase space is the inverse coherence
length $\kappa$. For a fixed composition, such as our
La$_{1.86}$Sr$_{0.14}$CuO$_{4}$ sample, $\alpha$ is fixed and
experiments are performed in the $T-\hbar\omega$ plane drawn.
Furthermore, $\alpha$ is associated with 
a particular inverse length  $\kappa_{o}$ when T and 
$\omega \rightarrow 0$. If
we add to the graphic description of the inset in Fig. 4 the
assumption of a Euclidean metric for measuring distances to the
QCP, we immediately recover Eq. 3 with the
dynamical critical exponent, $Z$=1. It turns out that theory for
2D quantum magnets supports the concept of the Euclidean
metric and hence that $Z$=1. In addition, it posits that $T$
and $\hbar\omega$ should be interchangeable, an idea labeled '$\omega/T$
scaling'\cite{ref17,ref7}. We have checked the extent to which our
data support these ideas by allowing $\kappa_{o}$, ${\rm Z}$, ${\rm E}
_ {\rm T}$, and ${\rm E}_{ \omega}$ all to vary to yield the best fit
of $\kappa(\omega,T)$. The outcome, namely that $\kappa _ o = 0.033
\pm 0.004$ \AA $^{ -1}$, ${\rm E} _ {\rm T} /{\rm k} _ {\rm B} = 590
\pm 100K$, ${\rm E} _
\omega /{\rm k} _ {\rm B} = 550 \pm
120K$ and ${\rm Z} = 1.0 \pm 0.2$, supports a simple QCP
hypothesis. 

Beyond providing a framework for understanding $\omega$- and T-dependent
length scales, the QCP hypothesis also has consequences for the
susceptibility amplitudes. Specifically, as  $\omega \rightarrow 0$, 
$\chi _ {\rm P} ^ {\prime \prime} ( \omega , {\rm T})/\omega$ should be
controlled by a single variable representing
the underlying magnetic length. In the upper right corner of Fig. 4, we plot 
$\chi _ {\rm P} ^{\prime\prime}/\omega$ as a function of such a variable, 
namely $\kappa(\omega=0,T)$. The outcome is 
that $\chi _ {\rm P} ^ {\prime \prime} ( \omega , {\rm T})/\omega$ 
is proportional to   $\kappa(\omega=0,T)^{\delta}$
where $\delta=(2-\eta+Z)/Z=3 \pm 0.3$, in agreement with 
theoretical expectations \cite{ref15} for
QCP's occurring in 2D insulating magnets.

To make the QCP hypothesis plausible, it would be useful to have
evidence for an ordered state nearby in phase space. Because the
high-T$_{c}$ superconductors can be chemically tuned, what we
are looking for are related compounds with magnetically ordered ground
states.  The most obvious is pure La$_{2}$CuO$_{4}$. However, beyond
the material itself seeming far away in the phase space of
Fig. \ref{schematics} {\bf A}, the simple unit cell doubling describing the
antiferromagnetism of the material is remote from the long-period spin
modulation which one would associate with the quartet of peaks seen in
the magnetic response of La$_{1.86}$Sr$_{0.14}$CuO$_{4}$. More
interesting compounds are found when the phase space is expanded to
consider ternary compounds,
where elements other or in addition to Sr are substituted onto the La
site.  When Nd is substituted
for La while keeping the Sr site occupancy(x) and hence hole density
at 1/8, the material is no longer superconducting but exhibits instead
a low-temperature phase characterized by magnetic Bragg peaks -
corresponding to static magnetic order- at loci close to where the
magnetic fluctuations are peaked in
La$_{1.86}$Sr$_{0.14}$CuO$_{4}$. While the full ternary phase diagram
has not been searched, we have sketched what it might look like in
Fig. 1, where the grey phase emerging close to the superconducting
state is the ordered 'striped phase' - so named because one model
describes it in terms of stripes of antiferromagnetic material
separated by lines of charges \cite{ref16}. More
generally, experiments on the high-T$_{c}$ materials can be thought of
as travels through a 3D phase space such as that
depicted in Fig. \ref{schematics} {\bf A}, and the changes in behavior
found on such travels can be associated with different features of the
landscape coming into prominence depending upon the height from which
they are observed. At the higher $\hbar \omega$'s and $T$s, the (red)
AF phase, characterized by a very high coupling
constant($\sim 0.15$ eV) is the most obvious feature. At the
intermediate $T$s we probed, the dominant feature
is the gray mountain where 'striped' order has been found. Finally, at
the lowest T, the superconducting instability dominates. The knowledge
that the cuprates inhabit an interesting 3D phase space together with
our discovery
that the spin fluctuations in one high-T$_{c}$ material 
are as singular
as the charge fluctuations should simplify the task 
of understanding both the anomalous normal state properties and
the high-Tc superconductivity of the cuprates.

\begin{figure}
\caption{{\bf A} Schematic phase diagram for La$_{2-x}$Sr$_x$CuO$_4$
showing the evolution from long range antiferromagnetic (AF) order 
(x=0, T$_N$ shown
in red), through an intermediate spin glass (SG) phase (green) to 
superconducting (SC)
order (blue).  Double doping (with Nd on the La site for example) results in 
stripe phase ordering (gray) shown along the y axis. {\bf B} Map of the region
of reciprocal space ({\bf Q} vector) near $(\pi,\pi)$ probed in the current 
measurements.  
Typically data were taken along the red line over two of the four 
incommensurate peaks that occur in our x=0.14 sample with the background 
determined along the dashed green trajectory.}
\label{schematics}
\end{figure}
\begin{figure}
\caption{{\bf A} Scans collected with unpolarized neutrons at constant energy ($\hbar \omega = 6.1$ meV) 
along the trajectory shown as a red line in Fig. 1 {\bf B}
(parametrized by $\xi$ where ${\bf Q}=\xi(\pi,\pi)+(\delta/2)(-\pi,\pi)$)
through two incommensurate peak positions.
Actual counting times were in the 10 to 60 minute per point
range.  Open symbols represent background collected along the trajectory
indicated by the dashed green line in Fig. 1 {\bf B}.
Solid lines correspond to resolution-corrected structure
factor defined by Eq. 1.  {\bf B} Polarized scans at constant energy
($\hbar \omega = 3.5$ meV)
at 40 K
showing the spin-flip and non spin-flip intensity.  The incommensurate
peaks occur only in the spin-flip channel confirming their magnetic
origin. {\bf C} and {\bf D} Energy and momentum (along solid trajectory
in Fig. 1 {\bf B}) -dependent magnetic response function 
$\chi ^ {\prime \prime} ({\rm Q}, \omega )$, derived from background
corrected intensities
using fluctuation-dissipation theorem at {\bf C} 35 and {\bf D} 297 K.
No attempt has been made to correct for experimental
resolution, which broadens and weakens sharp features in 
$\chi ^ {\prime \prime} ({\rm Q}, \omega )$.
Color scale corresponds to the raw background corrected intensities,
measured per unit signal in
the incident beam monitor,
while the (vertical) numerical scales are in units of 
counts per six minutes divided by [n($\omega$) + 1].  
{\bf E} Resolution-corrected (incommensurate) peak values
of the magnetic response as a function of frequency.}
\label{survey}
\end{figure}
\begin{figure}
\caption{Temperature dependence of {\bf A} peak intensity
derived from full polarization analysis\protect\cite{moon} and 
unpolarized neutron data at 3.5 meV and 
{\bf B} Resolution-corrected peak response divided by frequency in the 
low-frequency limit obtained from the fits described in the text. 
Absolute scale in {\bf B} is from normalization to
phonons\protect\cite{ref12}. {\bf C} inverse length scale 
$\kappa ( \omega , T)$ at various fixed energy transfers 
$\hbar \omega$.}
\label{tdep}
\end{figure}
\begin{figure}
\caption{Temperature dependence of inverse length scale 
$\kappa ( \omega , {\rm T})$ at various fixed energy transfers 
$\hbar \omega$ plotted against
T and $\hbar \omega$ added in quadrature.
The solid line corresponds to a Z = 1
quantum critical point (see Eq. 3 and text).
The graph in the upper right shows how the
peak response depends on $\kappa$ = $\kappa ( \omega=0, T)$.
The inset shows the three-dimensional space defined by $\omega$,T, and
a composition dependent control parameter $\alpha$. The dark
plane corresponds to the $( \omega , {\rm T})$ phase space
probed by our x=0.14 sample while the solid circle represents
a nearby quantum critical point.}
\end{figure}


\begin{references}
\bibitem{ref12} We have used acoustic phonons in scans along 
${\rm Q} = (2, \xi , 0)$ and (2,0, $\nu$ ) (orthorhombic notation)
for 
$\hbar \omega = 2$ and 2.7 meV (sound velocity 
$\cong 23 {\rm meV} -$ \AA).
\end{references}
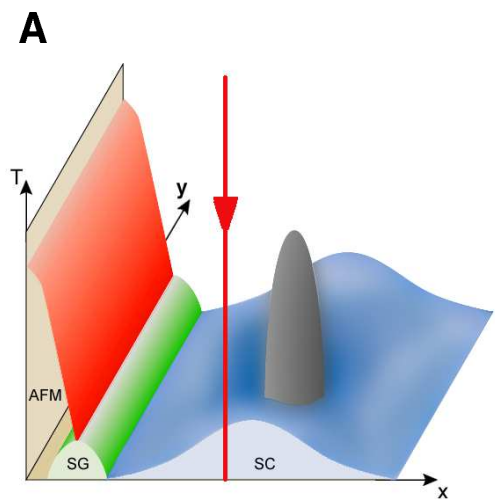
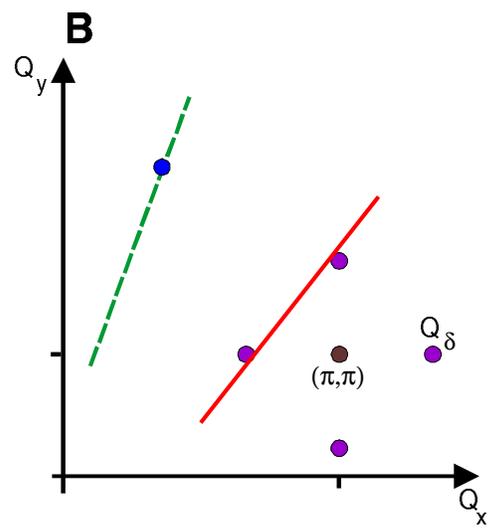
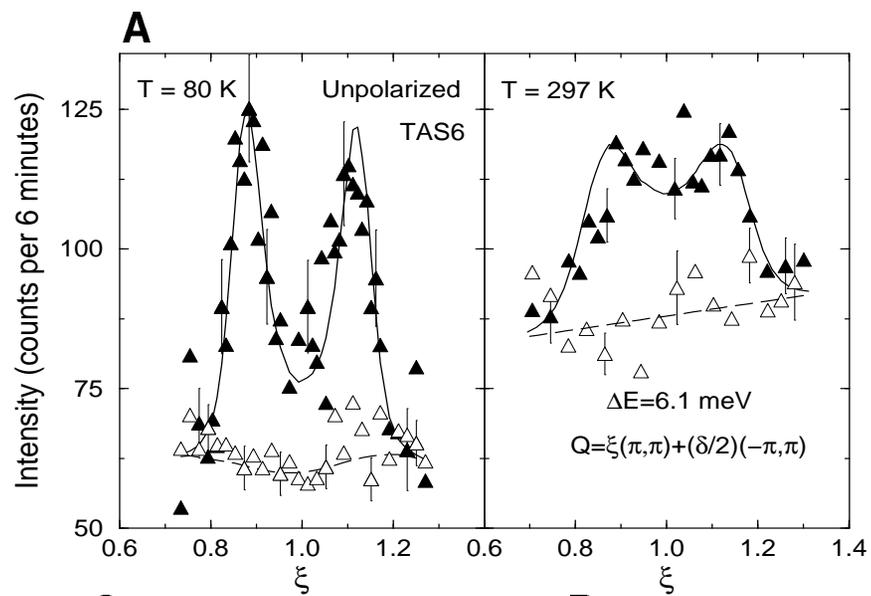
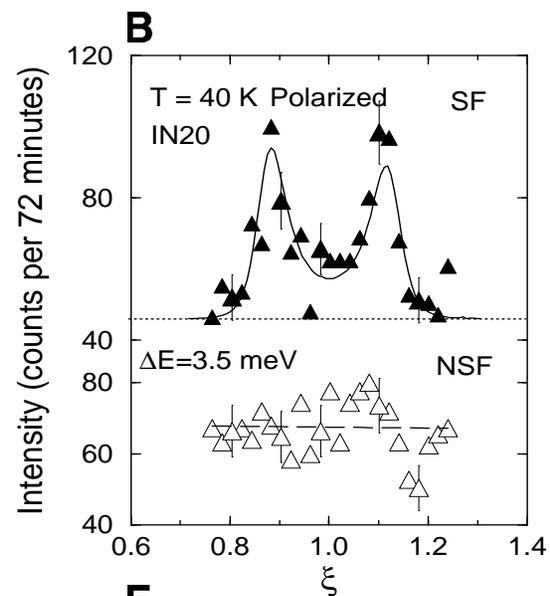
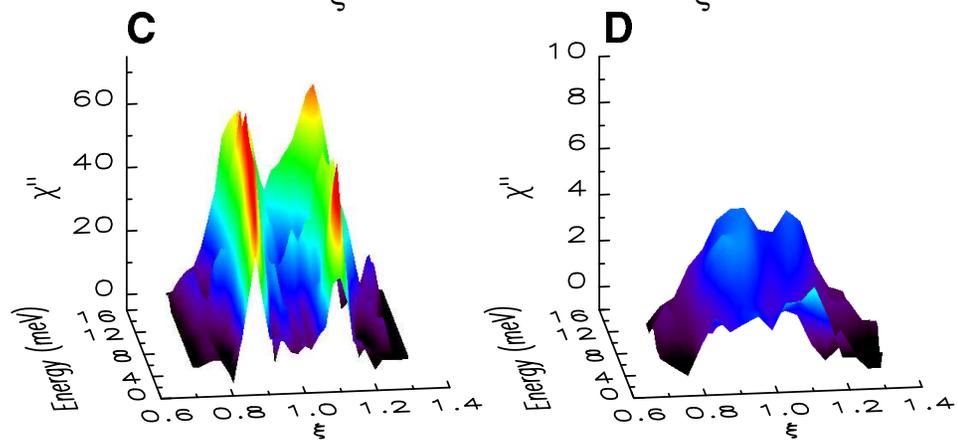
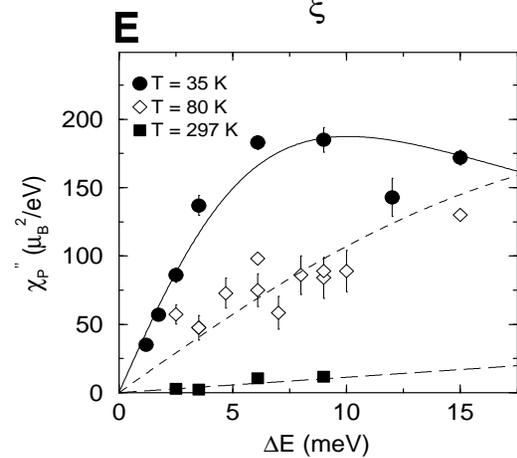
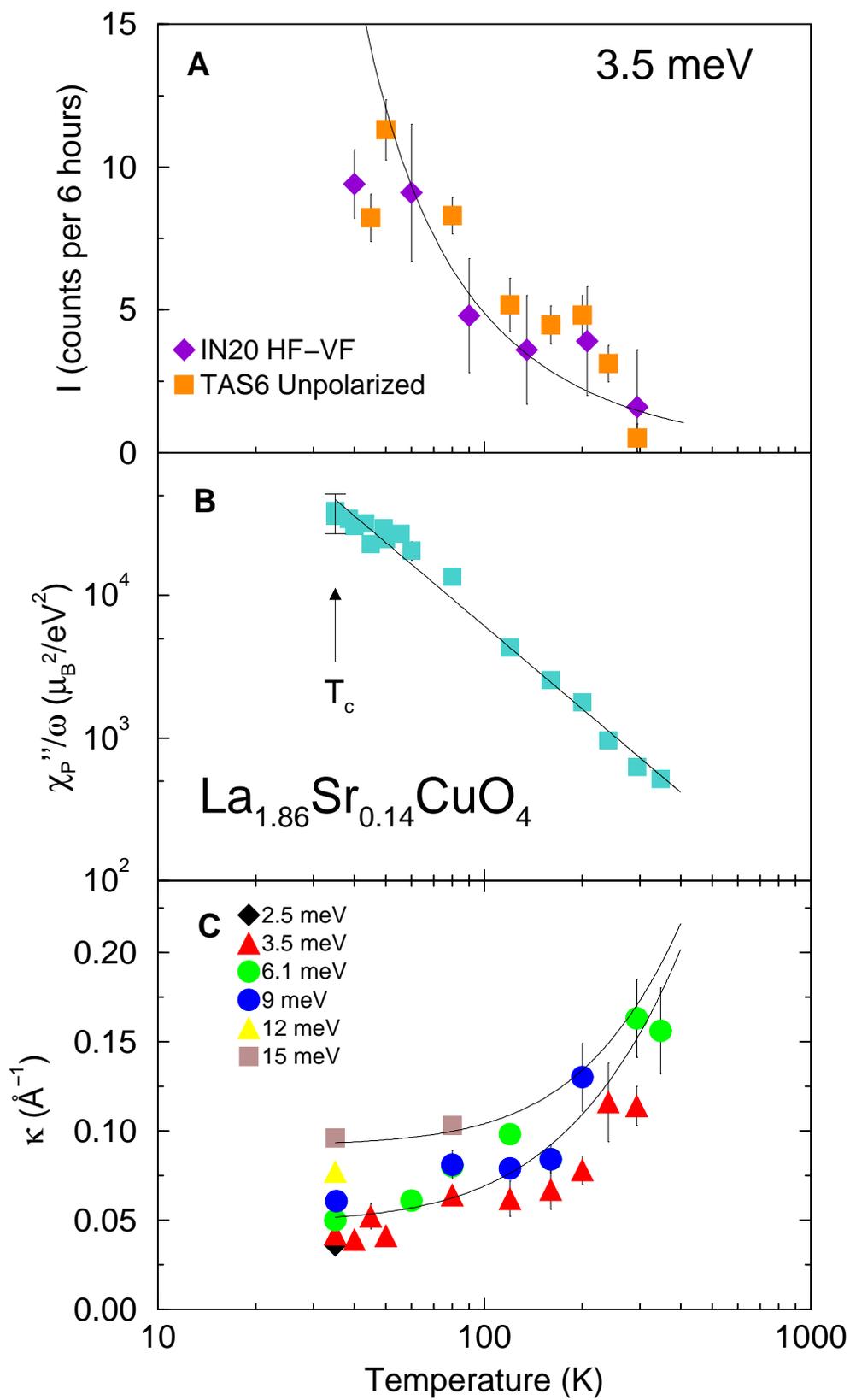
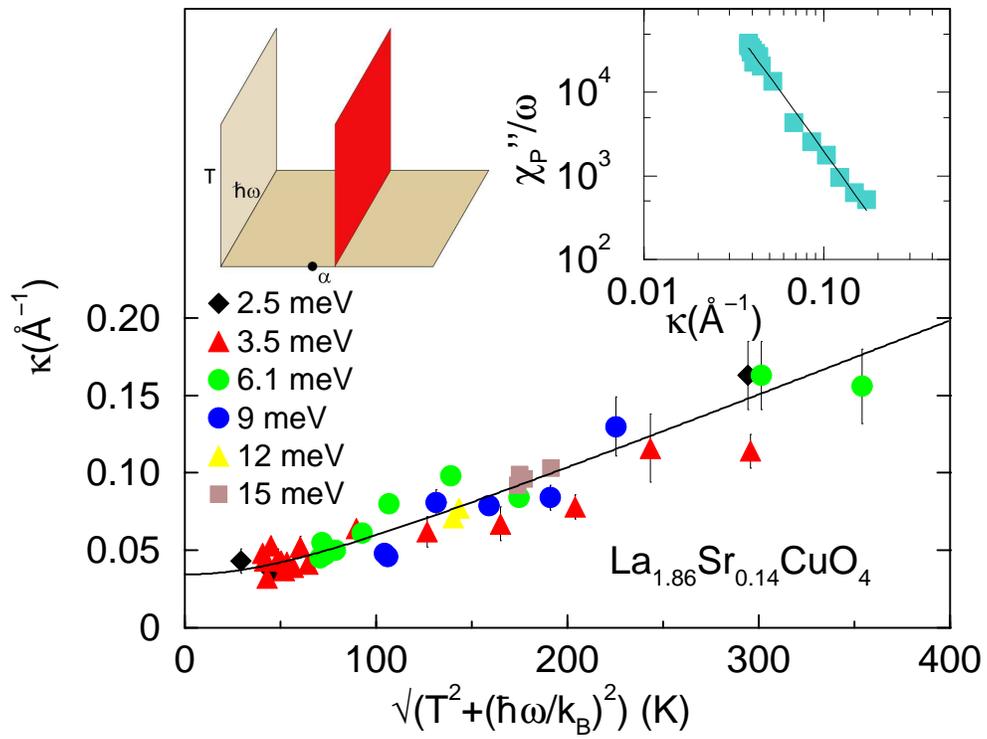
\end{document}